\documentclass[superscriptaddress,showpacs,aps,twocolumn,prl,floatfix,amssymb,amsmath]{revtex4-1}
\usepackage[dvips]{graphicx}
\usepackage{bm}
\usepackage{times}

\newcommand{\smeq}{\! = \!}

\newcommand{\smpl}{\! + \!}
\newcommand{\smmi}{\! - \!}

\newcommand{\bea}{\begin{eqnarray}}
\newcommand{\eea}{\end{eqnarray}}

\def\beq{\begin{equation}}
\def\eeq{\end{equation}}

\begin{document}

\title{Electrical Control of the Chemical Bonding of Fluorine on Graphene}
\author{J. O. Sofo}
\affiliation{Physics Department, Pennsylvania State University, University Park, PA16802,USA}
\author{A. M. Suarez}
\affiliation{Physics Department, Pennsylvania State University, University Park, PA16802,USA}
\author{Gonzalo Usaj}
\affiliation{Centro At{\'{o}}mico Bariloche and Instituto Balseiro,CNEA, 8400 Bariloche, and CONICET, Argentina}
\author{P. S. Cornaglia}
\affiliation{Centro At{\'{o}}mico Bariloche and Instituto Balseiro,CNEA, 8400 Bariloche, and CONICET, Argentina}
\author{A. D. Hern\'andez-Nieves}
\affiliation{Centro At{\'{o}}mico Bariloche and Instituto Balseiro,CNEA, 8400 Bariloche, and CONICET, Argentina}
\affiliation{Department of Physics, University of Antwerp, Groenenborgerlaan 171, B-2020 Antwerpen, Belgium}
\author{C. A. Balseiro}
\affiliation{Centro At{\'{o}}mico Bariloche and Instituto Balseiro,CNEA, 8400 Bariloche, and CONICET, Argentina}
\date{January 24, 2011.}

\begin{abstract}
We study the electronic structure of diluted F atoms 
chemisorbed on graphene using density functional theory calculations. 
We show that the nature of the chemical bonding of a F atom adsorbed 
on top of a C atom in graphene strongly depends on carrier doping.
In neutral samples the F impurities induce a $sp^3$-like bonding of
the C atom below, generating a local distortion of the hexagonal lattice.
As the graphene is electron-doped, the C atom retracts back to the graphene
plane and for high doping ($10^{14}$cm$^{-2}$) its electronic structure corresponds to a nearly pure $sp^2$ configuration. We interpret this $sp^3$-$sp^2$ doping-induced crossover in terms of a simple tight binding model and discuss the physical consequences of this change.
\end{abstract}
\pacs{73.22.Pr,81.05.ue,73.20.-r,73.20.Hb}
\maketitle

The unusual electronic and mechanical properties of graphene, which make it an excellent candidate to build nanoscale electronic devices, have generated an intense and exciting activity during the last few years 
~\cite{Novoselov2004,*Novoselov2005,Katsnelson2006,*Novoselov2007,*Geim2007,CastroNeto-review,*Beenakker2008,*DasSarma2010}. 
This two-dimensional (2D) honeycomb lattice of C atoms is a prototype 
of the C-C $sp^2$ binding. The cohesion and structural
stability of the lattice are mainly due to the $\sigma $-bonds, while the $p_{z}$-orbitals  
determine
its low energy electronic properties. 
The anomalous charge transport in pure graphene is a consequence of the chiral nature of the low energy
excitations, which correspond to massless Dirac fermions~\cite{CastroNeto-review,*Beenakker2008}, and its linear dispersion relation close to the Fermi energy. This leads to a strong suppression of the backscattering processes which, together with the high purity inherent to the samples, is responsible for the long electron mean-free path.
 Clean graphene samples can be grown because substitutional impurities, vacancies and other structural defects are quite rare due to the stability of  the strong $\sigma $-bonds. 
Therefore, one of the main sources of electron scattering is the presence of adsorbed impurities. For this reason, the control and manipulation of adatoms is viewed as a possible route to modify the electronic properties and to create the appropriate conditions for charge and spin transport.

The problem of adatoms on graphene has been theoretically and experimentally
studied by several groups~\cite%
{Lehtinen2003,Duplock2004,Meyer2008,Chan2008,Uchoa2008,*CastroNeto2009a,Cornaglia2009,Wehling2009,*Wehling2010,*Wehling2010a,CastroNeto2009,Boukhvalov2009,Ao2010}%
. It is known that impurities like H or F are
adsorbed on top of a C atom---from hereupon denoted as C$_{0}$---and
induce a distortion of the flat honeycomb lattice. This distortion
consists of a puckering of the C$_{0}$ carbon as its electronic configuration
changes from $sp^2$ to another with a more $sp^3$ character (see Fig.\ref%
{adatom}). The behavior of H and F impurities on neutral graphene
presents some quantitative and qualitative differences: on the one hand, the charge transfer
is larger and the graphene distortion smaller for F impurities. On the other hand, there are
strong indications that H creates a magnetic defect while no magnetism is
obtained with F. However, spin related effects are to be expected even in the case of F impurities as the local distortion of the $2$D plane leads, like in carbon nanotubes~\cite{DeMartino2002,Huertas-Hern2006}, to spin-orbit coupling ~\cite{CastroNeto2009} .

\begin{figure}[t]
\centering
\includegraphics[width=.12\textwidth]{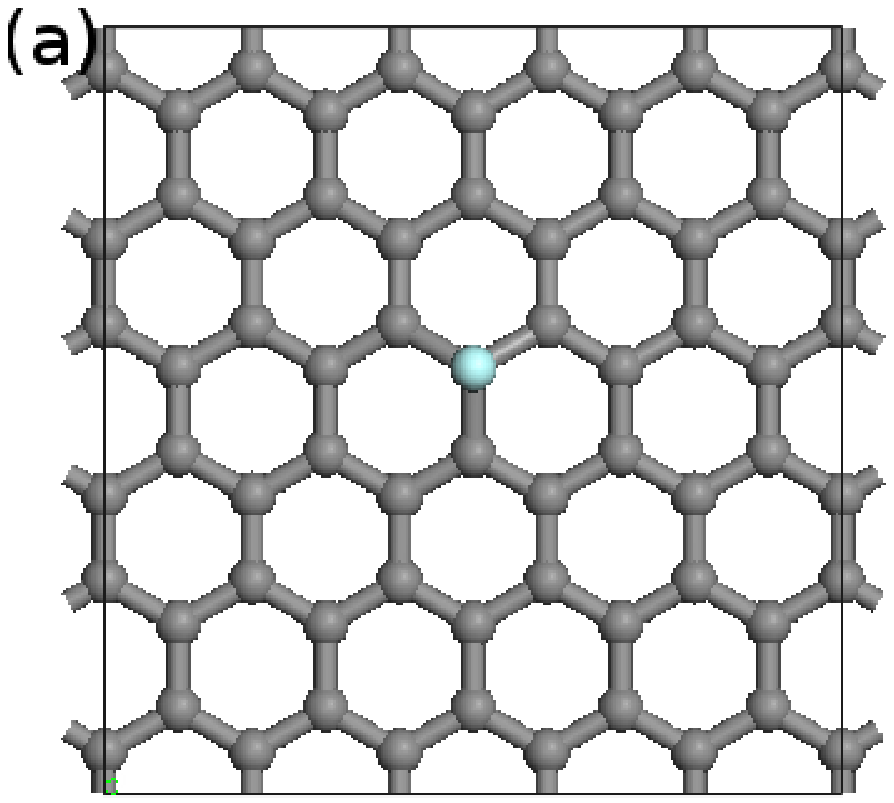}
\includegraphics[width=.12\textwidth,clip]{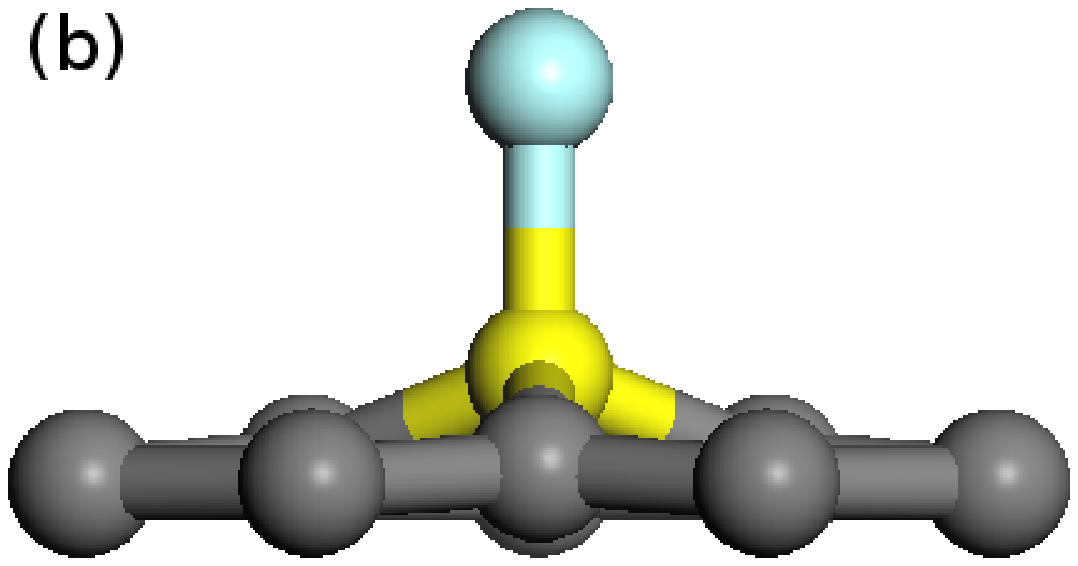} 
\includegraphics[width=.45\textwidth]{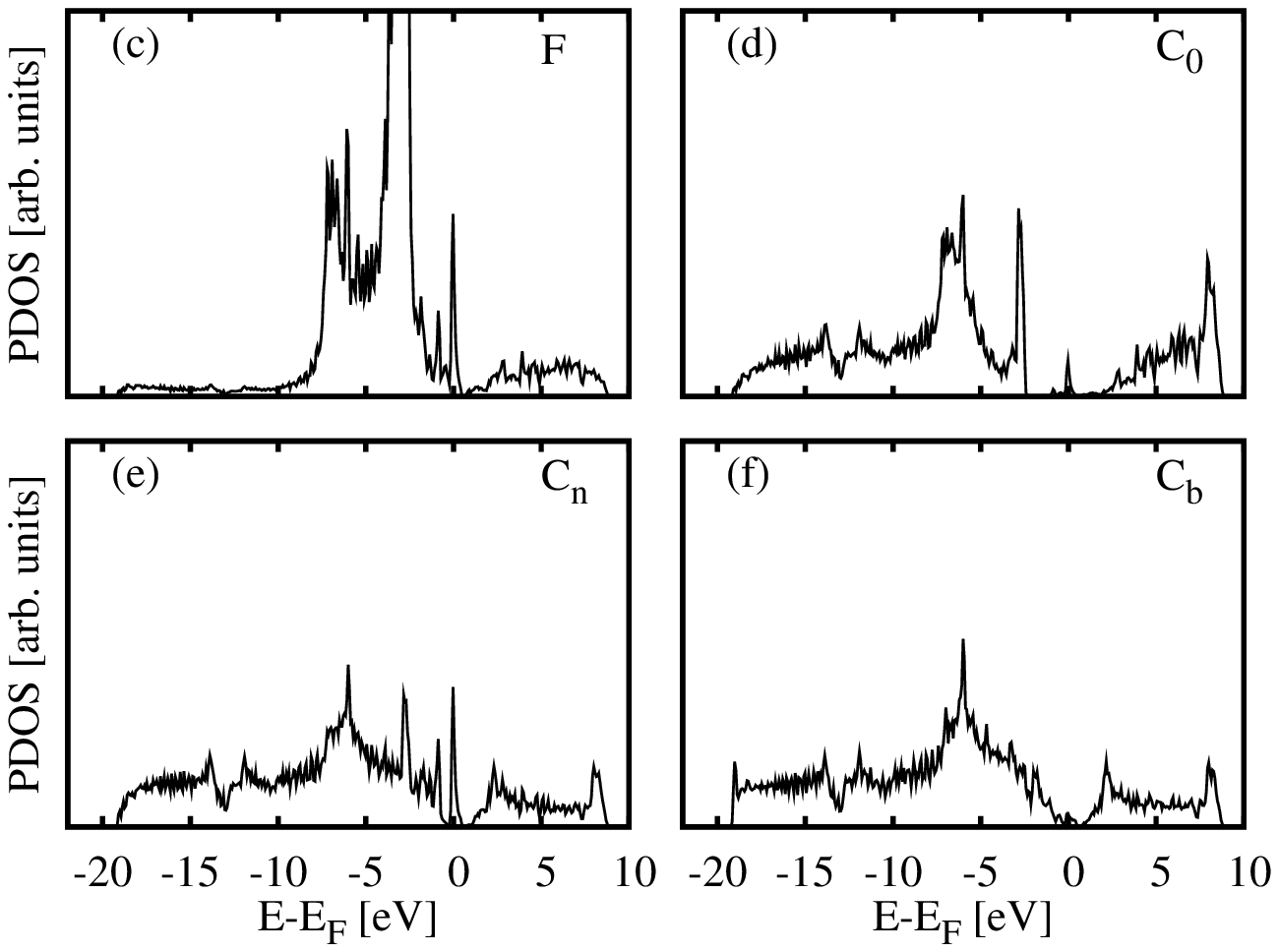}
\caption{(a) Unit cell used in our DFT calculations (60 C atoms); (b) schematic lateral view of the defect showing the local puckering of the graphene sheet caused by the adatom; Partial density of states (PDOS) for the undoped case projected on the F (c), the C$_0$ atom (d), a C$_0$-nearest neighbor C atom, C$_n$ (e), and a `bulk' C$_b$ atom (f). 
}
\label{adatom}
\end{figure}

Despite this intense activity, little is known about the effect of graphene's doping on the character of the bonding of atoms and molecules. Here, we show that carrier doping determines the 
structure of the defect induced by F impurities and the nature of the chemical bond. 
With electron-doping, the distortion of the lattice decreases and the F-C$_{0}$
distance increases. Notably, for large enough doping, the F-C bonding acquires an ionic character. 
This effect provides an opportunity to study, in a
controllable and continuous way, the evolution from ionic-like ($sp^2$
configuration of  C$_{0}$) to covalent-like ($sp^3$
configuration of C$_{0}$) bonding of adatoms on graphene. Hole-doping, as opposed to electron doping, results in a moderate increase of the lattice distortion.
These changes have important consequences on the local spectroscopic properties of the adatom and its surrounding atoms and will certainly modify  the charge and spin transport properties of the system.

\begin{figure}[t]
\centering
\includegraphics[width=.45\textwidth]{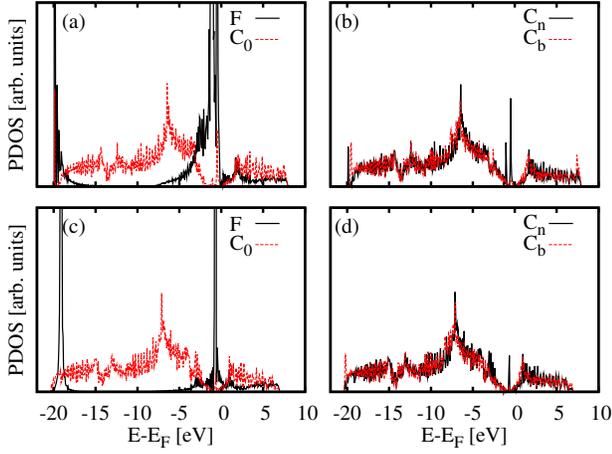}
\caption{PDOS for electron doped graphene with one (a,b) and two (c,d) additional electrons per DFT unit cell. Labels are the same as in Fig. \ref{adatom}. Notice the reduction of the hybridization between the F atom and the graphene sheet as the doping increases.}
\label{DOS-doping-total}
\end{figure}

We consider a graphene layer with a unit cell of $60$ C atoms and one F atom (a $1.6\%$ concentration of F atoms). 
A schematic side view of the distortion induced by the F atom is shown in Fig. \ref{adatom}(b).
Figures \ref{adatom}(c) and \ref{adatom}(d) show the partial density of states (PDOS) of an
undoped graphene layer obtained using Density Functional Theory (DFT)  
\footnote{DFT calculations were done with a plane wave basis as implemented in the VASP code \cite{Kresse1996a,*Kresse1996} and the cutoff energy set to $400$ eV. The core electrons were treated with a frozen core
projector augmented wave (PAW) method \cite{Blochl1994,*Kresse1999}. F and C's $1s$ electrons were treated in
the core. The projector radius for F was $1.19$\AA{} while for C was $1.20$\AA. We use the PBE
generalized gradient approximation to treat the exchange and correlation \cite{Perdew1996,*Perdew1997}. To correct for the
dipole moment generated in the cell and to improve convergence with respect to the periodic cell
size, monopole and dipole corrections were considered \cite{Neugebauer1992,*Makov1995}.}.
These results make evident that there is a large hybridization between the F 
and the C$_0$ atom---the strong broadening of the F's atomic levels is consistent with a covalent bonding. 
We also note that there is a narrow resonance close to the Fermi energy, similar to what is obtained for H. However, in the case of F, as opposed to the H case, this vacancy-like state does not lead to the appearance of magnetic order~\cite{Usaj2010}.

For a quantitative analysis of the distortion, we follow Ref. [%
\onlinecite{CastroNeto2009}] and relate the angles between the bonds with
the $sp$-character of the C$_{0}$ orbitals. In terms of its $s$ and $%
p_{z} $ atomic orbitals, the hybrid $\pi $-orbital
can be written as 
%\begin{equation}
$| \pi \rangle \smeq A| s \rangle+\sqrt{1-A^{2}}| p_{z} \rangle  \label{piket}$
%\end{equation}%
where $A\smeq 0$ and $A\smeq \frac{1}{2}$ correspond to the $%
sp^2$ and $sp^3$ configurations, respectively. A local basis includes the $%
\sigma $-orbitals that are also parameterized by the same constant $A$ ~\cite%
{CastroNeto2009}. The angle $\theta $ between the $\sigma $-bonds and the $z$%
-axis, perpendicular to the graphene plane, is given by $\cos {\theta }\smeq -A/%
\sqrt{2+A^{2}}$. This expression interpolates between $\theta \smeq 90^{\circ }$
for the $sp^2$ and $\theta \smeq 109.47^{\circ }$ for the $sp^3$ configurations.
Our results for the case of diluted F atoms on neutral graphene give $%
\theta \smeq 102.5^{\circ }$ and $A\smeq 0.314$. These values indicates a dominant $sp^3$
character of the bonds of the C$_{0}$ carbon atom. They are in qualitative
agreement with previous results obtained with a high F concentration but are smaller than the distortion found in F-graphene~\cite{Zhou2010}.

\begin{table}[t]
\begin{ruledtabular}
\begin{tabular}{lllll}
Electrons & $\phi$ & $\theta$ & $d_{\mathrm{F-C}_0}$ [\AA{}] & $d_{\mathrm{C}_0\mathrm{-C}_n}$ [\AA{}]\\
$-2$ & $115.1$ & $103.1$ & $1.47$ & $1.49$\\
neutral & $115.5$ & $102.5$ & $1.57$ & $1.48$\\
$+1$ & $118.8$ & $96.3$ & $2.00$ & $1.43$\\
$+2$ & $119.9$ & $91.9$ & $2.50$ & $1.42$
 \end{tabular}
 \end{ruledtabular}
\caption{Impurity induced graphene distortion for different numbers of
electrons added (or removed) from the unit cell. Here, $\protect\phi$ is the
angle formed by the C$_0$ atom and two of its nearest-neighbors (C$_n$); $%
\protect\theta$ is the angle of the C$_0$-C$_n$ $\sigma$-bonds and the normal to
the graphene plane; $d_{\mathrm{F-C}_0}$ and $d_{\mathrm{C}_0\mathrm{-C}_n}$ are
the F-C$_0$ and C$_0$-C$_n$ bond lengths, respectively.}
\end{table}

To simulate electron and hole doping we add or remove electrons from the
unit cell. The extra charge is compensated with a uniform charged
background. The results obtained by adding one and two electrons per DFT unit cell,
corresponding to dopings of the order of $\sim\!10^{14}$cm$^{-2}$, are
shown in Fig. \ref{DOS-doping-total}. The F's PDOS shows an increasing narrowing of the resonances as the doping is increased while, at the same time, the C$_{0}$'s PDOS becomes more similar to the one corresponding to a `bulk' carbon atom (C$_{\mathrm{b}}$). This indicates an important reduction of the hybridization parameter $V$ as a consequence of the new structure of
the defect in the doped system. Namely, with electron-doping the C$_{0}$ carbon atom
retracts back to the graphene plane and the F-C$_{0}$ distance increases.
This change is accompanied by an increase of the charge transfer to the F atom. For the largest doping studied here, two added electrons per DFT unit cell, we obtain $\theta \smeq 91.9^{\circ }$ which gives $A\smeq 0.047$. These values
correspond to an almost pure $sp^2$ configuration of C$_{0}$. On the
contrary, with hole-doping, the distortion increases and the F-C$_{0}$
distance decreases, indicating an increase of the $sp^3$-character of the bonding of the C$_{0}$
carbon atom---see Table I for details.

Hence, our results indicate that the effect of the F adatom on graphene can be strongly modified by external gates, a simple way to change doping, offering a new route to control graphene's transport properties. For instance, as mentioned above, the local distortion of the lattice introduces a spin-orbit coupling. As discussed in Ref. \cite{CastroNeto2009}, the effective  spin-orbit coupling is roughly given  by 
$\Delta _{so}\smeq -$ $\Delta_{so}^{at}A\sqrt{3(1-A^{2})}$, where $\Delta_{so}^{at}$ is the atomic spin-orbit coupling of the C atoms.
Then, our results show that, as the system is doped with electrons, $\Delta_{so}$ changes by a factor  $\sim6$.
Such a change should be clearly detected in spin dependent transport measurements.

\begin{figure}[t]
\centering
\includegraphics[width=.45\textwidth]{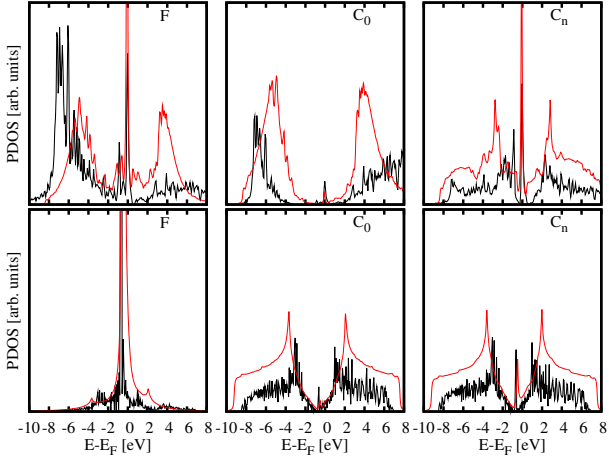}
\caption{Partial density of states projected onto the $p_z$ atomic orbitals for the F, C$_0$, and C$_n$ atoms for undoped graphene (top panels) and for the electron doped ($+2$) case (bottom panels). The thick lines correspond to the DFT results while the thin lines are the results of the tight binding model with the parameters discussed in the text.The unit cell contains $60$ C atoms in both calculations.}
\label{DOS-doping-pz}
\end{figure}

In order to understand the physics underlying this $sp^3$-$sp^2$ crossover we
use a minimal tight binding (TB) model that captures the relevant features. The $%
\pi $-bands of the graphene plane are described by a Hubbard Hamiltonian 
%\begin{equation}
$H_{\mathrm{graph}}\smeq-\sum_{\langle i,j\rangle \sigma }t_{ij}\,c_{i\sigma
}^{\dag }c_{j\sigma }^{}\smpl U_{c}\sum_{i}\,\left( \hat{n}_{i\uparrow }\smmi\frac{1}{2}%
\right) \left( \hat{n}_{i\downarrow }\smmi\frac{1}{2}\right) ,
$
%\end{equation}%
where $c_{i\sigma }^{\dag }$ creates an electron with spin $\sigma $ at site 
$i$ of the graphene lattice, $\hat{n}_{i\sigma }\smeq c_{i\sigma }^{\dag }c_{i\sigma }^{{}}$, and the first sum runs over nearest neighbors. The F
impurity is bounded to the C$_{0}$ atom, located at site $i\smeq0$ and
represented by the relevant $|p_z \rangle $ orbital, and it is described by 
%\begin{equation}
$H_{f}\smeq\sum_{\sigma }\varepsilon _{f}\,f_{\sigma }^{\dag }f_{\sigma
}^{}+U_{f}\,f_{\uparrow }^{\dag }f_{\uparrow }^{}f_{\downarrow }^{\dag
}f_{\downarrow }^{}$,
%\end{equation}%
where $f_{\sigma }^{\dag }$ creates an electron with spin $\sigma $ and
energy $\varepsilon_{f}$ at the relevant orbital of the impurity and $U_{f}$
is the intra-atomic Coulomb repulsion. The impurity-graphene interaction
includes a one-body hybridization $V$, a distortion-induced shift of the
C$_{0}$ energy $\Delta $, the interaction with the image charge $V_{\mathrm{im}}$ ~\cite{NEWNS1969} and a local inter-atomic interaction $U_{f0}$, 
\bea
\nonumber
H_{\mathrm{int}}&\smeq &\Delta \,\hat{n}_{0}-V\,\sum_{\sigma }(c_{0\sigma
}^{\dag }f_{\sigma }^{}\!+\!f_{\sigma }^{\dag }c_{0\sigma }^{})\\
&&-V_{\mathrm{im}}\,(\hat{n}_{f}-1)^{2}+U_{f0} (\hat{n}_f-1)(\hat{n}_0-1),
\eea
with $\hat{n}_{0}\smeq c_{0\uparrow }^{\dag}c_{0\uparrow }^{{}}\!+\!c_{0\downarrow }^{\dag }c_{0\downarrow }^{{}}$ and $\hat{n}_{f}\smeq f_{\uparrow }^{\dag }f_{\uparrow }^{{}}\!+\!f_{\downarrow}^{\dag }f_{\downarrow }^{{}}$. 
Note that the effect of $V_\mathrm{im}$ is to renormalize both the orbital energy, $\tilde{\varepsilon}_f\smeq \varepsilon _f+V_\mathrm{im}$, and the intra-atomic repulsion, $\tilde{U}_f\smeq U_f-2V_\mathrm{im}$, favoring the charging of the F atom. The $U_{f0}$ term, instead, takes into account the discharge induced on the C$_0$ atom upon charging of the F atom.
In what follows, guided by the DFT calculations, we take  $t_{ij}$ with $%
i,j\neq 0$ equal to $t\smeq 2.8eV$ and include the reduction of $t_{0j}\smeq t_{j0}$ induced by the distortion in the $sp^3$ case. The energy shift of C$_0$  is given by $\Delta \simeq A^{2}(\varepsilon _{s}-\varepsilon _{p})$~\cite{CastroNeto2009}, where $\varepsilon _{s}\!\sim\! -8$eV and $\varepsilon_{p}\!\equiv\!0$ are the energies of the $s$ and $p$ carbon orbitals,
respectively.
Since the parameters $t_{0j}$, $\Delta $ , $V$, $V_\mathrm{im}$ and $U_{f0}$ depend on the distortion of the
lattice, and to minimize the number of free parameters of the model, we
consider two frozen configurations corresponding to the two limiting cases
presented above: \textit{i}) the distorted lattice of Fig. \ref{adatom}
with $\Delta \smeq -0.3t$, $t_{0j}\smeq 0.7t$,  and $V\smeq 1.4t$~\cite{Ihnatsenka2010}; \textit{ii}) the undistorted lattice with $\Delta \smeq 0$, $t_{0j}\smeq t$  and $V\smeq 0.4t$ ---the smaller value of the hybridization $V$ reflects the larger F-C$_{0}$ distance shown in Table I. In the latter case, where the graphene sheet is essentially unalterated, we can consider that the situation is reminiscent of the classical problem of a point charge in front of a conducting plate and take, as a simple approximation, $V_{\mathrm{im}}^{sp^2}\smeq\gamma e^{2}/4d_{\mathrm{F}-\mathrm{C}_0}$ with $\gamma\!\sim\!0.8$~\cite{NEWNS1969,Ghaznavi2010}. On the contrary, in the $sp^3$ case, the F is covalently bounded to C$_0$ and then one expects a stronger local screening of the F's excess charge. That is, we expect the term $U_{f0}$ to be dominant. For the sake of simplicity, and to avoid a double counting of the Coulomb interaction,  we take  $V_{\mathrm{im}}^{sp^3}\smeq  0$ and choose $U_{f0}^{sp^3}\!\simeq\!V_{\mathrm{im}}^{sp^2}$. Similarly, we take  $U_{f0}^{sp^2}\smeq0$.
In the following, we solved the Hamiltonian $H\smeq H_{\mathrm{graph}}+H_f+H_{\mathrm{int}}$ in
the Hartree-Fock approximation. From comparison to the DFT results for the PDOS of the F, C$_0$ and C$_n$ atoms  we estimate the effective parameters $U_{f}\!\sim \!2.5t$ and $\varepsilon_{f}\!\sim\!-1.8t$ while we set $U_{c}\smeq t$.

\begin{figure}[t]
\centering
\includegraphics[width=0.4\textwidth]{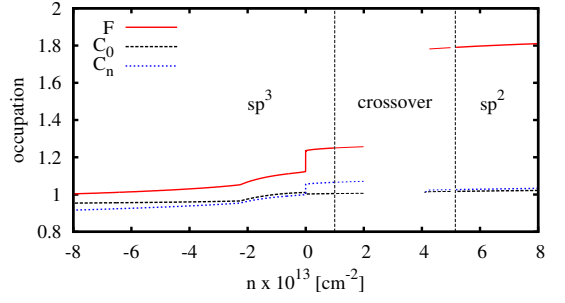}
\caption{Occupation of the F, C$_0$, and C$_n$ atoms for a simplified tight binding model in the $sp^3$ and  $sp^2$ configurations, as a function of the carrier density. Note that the charge of the F atom is higher in the $sp^2$ configuration.}
\label{HF-results}
\end{figure}

Figure \ref{DOS-doping-pz} shows a comparison of the DFT and TB results for the PDOS projected onto the $p_z$ orbitals. While the latter is not expected to give an accurate description of the former, there is a good qualitative agreement between both methods for the general features of the PDOS, which gives support to the chosen effective TB parameters. 
Figures \ref{HF-results}(a) and %
\ref{HF-results}(b) show the occupation of the F, C$_0$ and
one of the C$_n$ atoms as a function of the graphene's electron density for
the $sp^3$ and $sp^2$ configurations, respectively. 
In the $sp^3$ case the F, C$_0$ and C$_{n}$ atoms are moderately (dis-)charged
as the graphene is doped with electrons (holes)---note that C$_{0}$ is slightly discharged when F is charged. This differs from the $sp^2$ case where the F
atom is heavily charged while the C$_{0}$ and C$_{n}$ atoms remain essentially uncharged. This is consistent with the DFT results, where a significant increase of the 
 charge of the F atom is observed in the electron doped case---in addition to that, the DFT shows a small discharge of the C$_0$ atom, which in our model can be obtained by including a small $U_{f0}^{sp^2}\!\simeq\!0.1t$. 

 It is worthy of mention that the  simplified model also helps to understand the absence of magnetism in the DFT calculations. This is so because the main features can be described by an even simpler model where interactions are only included on the F, C$_0$ and C$_n$ atoms. In that case, the system behaves as an effective Anderson impurity and then the absence of magnetism can be understood in terms of its effective parameters~\cite{Usaj2010}.

The mechanism behind this covalent-ionic crossover is the competition between the hydridization of the F and C$_0$ atoms and both the distortion of the lattice and the Coulomb interaction. In the neutral case, the $sp^3$ configuration is more stable since the energy gain due to F-C$_0$ hybridization overcomes the energy lost due to the lattice distortion. 
From DFT calculations, we estimate the latter to be $\!\sim\!0.4t$. Since it mainly comes from the  $\sigma$ bands this contribution is not included in our TB model---we assume it to be doping independent.
Upon doping, the increase in the occupation of both the F and $C_0$ (enhanced in the latter case by the energy shift $\Delta$) increases the energy due to the local Coulomb interaction $U_{f0}$ while, at the same time, it leads to a reduction of the gain due to $V$ (`antibonding' states become occupied). On the contrary, the $sp^2ˆ$ configuration, which does not pay the distortion energy, gains some energy due to $V_\mathrm{im}$ when the F is charged. 
To verify this scenario, we calculate the energy difference
between the two configurations~\cite{NEWNS1969}, 
$
\Delta E\smeq E_{sp^2}\smmi E_{sp^3}\smeq\int \langle \frac{\partial \hat{H}}{\partial
\lambda }\rangle \,\mathrm{d}\lambda
$, 
where $\lambda $ is a parameter that linearly interpolates all the
parameters that change between the two configurations ($\Delta $, $t_{oj}$, $%
V$, $V_{\mathrm{im}}$ and $U_{f0}$) and  $\langle \dots \rangle $ is the average
value in the Hartree-Fock state.
Within our TB model, we found that $\Delta E\sim 0.4t$ is roughly constant for hole doping but it drops as the graphene sheet is doped with electrons.  
While this Hartree-Fock result is sensitive to the chosen parameters and even though this is an oversimplified picture of the real process where the local configuration is
progressively distorted by the doping (see Table I), it clearly gives further support to our interpretation the DFT results.

In summary we have shown that the nature of the F-C$_{0}$
bonding changes with doping from  covalent for hole-doping to  ionic  for
electron-doping. This is followed by a change of the structure of the
defect ($sp^3\!\rightarrow\! sp^2$) with an important distortion of the lattice for the hole-doped system
to an almost undistorted lattice for electron-doped graphene. This structural change can serve as a new knob to control graphene's electronic properties. As mentioned before,  the properties of several elements and molecules on neutral  graphene have been extensively studied~\cite{Chan2008,Wehling2009}. Our results point out the relevance of doping and call for a more exhaustive analysis of other elements, beyond F, as they could present similar effects.

JOS and AS acknowledge support from the Donors of the American Chemical Society Petroleum Research Fund and use of facilities at the Penn State Materials Simulation Center. GU, PSC, ADH, and CAB acknowledge financial support from PICTs 06-483 and 2008-2236 from ANPCyT and
PIP 11220080101821 from CONICET, Argentina. ADH acknowledges support from the Flemish Science Foundation (FWO).

%\bibliographystyle{apsrev4-1}
%\bibliography{../../graphene}
%merlin.mbs apsrev4-1.bst 2010-07-25 4.21a (PWD, AO, DPC) hacked
%Control: key (0)
%Control: author (72) initials jnrlst
%Control: editor formatted (1) identically to author
%Control: production of article title (-1) disabled
%Control: page (0) single
%Control: year (1) truncated
%Control: production of eprint (0) enabled
%

\end{document}